\documentclass[10pt, apj]{emulateapj}

\usepackage{graphicx} 
\usepackage{natbib} 
\usepackage{amsmath}
\usepackage{amssymb} 
\usepackage{fancyhdr} 
\usepackage{latexsym}
\usepackage{color} 
\usepackage{txfonts}
\definecolor{mypurple}{RGB}{128,0,255}
\usepackage[hyperfootnotes=false, colorlinks = true, linkcolor = blue,
  urlcolor = blue, citecolor=mypurple]{hyperref}

\graphicspath{{./}}
\usepackage{soul}
\usepackage{aas_macros}
\usepackage{epstopdf}

\newcommand\asd[1]{{\color{black} #1}}
\newcommand\asdrev[1]{{\color{black} #1}}
\newcommand\asdrevok[1]{{\color{black} #1}}

\usepackage{mydefs_iris2}

\newcommand\bdp[1]{{\color{black} #1}}

\newcommand{\longacknowledgment}{This work is supported by NASA under contract
NNG09FA40C ({\it IRIS}) and the Lockheed Martin Independent Research Program. JdlCR is supported by grants from the Swedish Research Council (2015-03994), the Swedish National Space Board (128/15) and the Swedish Civil Contingencies Agency (MSB). This project has received funding from the European Research Council (ERC) under the European Union's Horizon 2020 research and innovation programme (SUNMAG, grant agreement 759548). \bdp{This paper has benefited from discussions at a meeting of team 399 ’Studying magnetic-field-regulated heating in the solar chromosphere’ at the International Space Science Institute (ISSI) in Switzerland. IRIS is a NASA small explorer mission developed and operated by LMSAL with mission operations executed at NASA Ames Research center and major contributions to downlink communications funded by ESA and the Norwegian Space Centre.}}

\begin{document}

\title{RECOVERING THERMODYNAMICS FROM SPECTRAL PROFILES OBSERVED BY IRIS: \\ A MACHINE AND DEEP LEARNING APPROACH}

\author{Alberto Sainz Dalda\altaffilmark{1}$^,$\altaffilmark{2}$^,$\altaffilmark{3}, Jaime de la Cruz Rodr{\'i}guez\altaffilmark{4}, Bart De Pontieu\altaffilmark{2,5,6}, and Milan Go\v{s}i\'{c}\altaffilmark{1}$^,$\altaffilmark{2}}
\altaffiltext{1}{Bay Area Environmental Research Institute, PO Box , CA, USA}
\altaffiltext{2}{Lockheed Martin Solar and Astrophysics Lab, Palo Alto, CA 94304, USA}
\altaffiltext{3}{Stanford University, Stanford, CA 94305, USA}
\altaffiltext{4}{Institute for Solar Physics, Dept. of Astronomy, Stockholm University, AlbaNova University Centre, SE-106 91 Stockholm, Sweden}
\altaffiltext{5}{Rosseland Centre for Solar Physics, University of Oslo, P.O. Box 1029 Blindern, N-0315 Oslo, Norway}
\altaffiltext{6}{Institute of Theoretical Astrophysics, University of Oslo, P.O. Box 1029 Blindern, N-0315 Oslo, Norway}

\submitted{}
\accepted{April 2, 2019}

\begin{abstract}
Inversion codes allow reconstructing a model atmosphere from observations. With the inclusion of optically thick lines that form \asd{in} the solar chromosphere, such modelling is computationally very expensive because a non-LTE evaluation of the radiation field is required.
In this study, we combine the results provided by these traditional methods with machine and deep learning techniques to obtain similar-quality results in an easy-to-use, much faster way. We have applied these new methods to \mgii lines observed by IRIS. As a result, we are able to reconstruct the thermodynamic state (temperature, line-of-sight velocity, non-thermal velocities, electron density, etc.) in the chromosphere and upper photosphere \asdrevok{of an area equivalent to an active region} in a few CPU minutes, speeding up the process by a factor of $10^5-10^6$. The open-source code accompanying this paper will allow the community to use IRIS observations to open a new window \bdp{to} a host of solar phenomena.
\end{abstract}


\section{Introduction}
To answer some of the major open questions about the solar atmosphere, 
it is critical to understand the physical conditions in the chromosphere.
The chromosphere has been observed for decades from ground- and space-based telescopes.
Particularly the {\it Interface Region Imaging Spectrograph}  explorer \citep[IRIS]{DePontieu14a} has observed 
more than $\approx19,000$ data sets at sub-arcsecond resolution in the \mgii spectral range, in the near ultraviolet (NUV), since it was launched \bdp{in} 2013. 

The formation of the \mgii lines has been studied using numerical calculations that include the effect of partial redistribution of scattered photons and 3D radiative transfer effects \citep{Leenaarts13b,Leenaarts13a,delPinoAleman16,Sukhorukov17}. Some spectral features such as the intensity and wavelength of the emission peaks and central reversal of those lines \bdp{can potentially serve} as proxies of the temperature, line-of-sight velocity ($v_{los}$), and their gradients in various regions of the chromosphere \citep{Leenaarts13a, Pereira15}. \bdp{However, so far
 these proxies have only been studied for quiet-Sun like conditions, and do not provide detailed height-dependent diagnostics.}

One of the most successful methods to recover physical information from spectropolarimetric observations is through non-linear fitting techniques, where the parameters of a model atmosphere are iteratively adjusted in order to match the \bdp{emerging model intensities with the observed spectra}. This procedure is commonly called an "inversion" even though it is not based on a formal inversion to the radiative transfer equation. 

\cite{delaCruzRodriguez16} and \cite{2018arXiv181008441D} have developed the {\it STockholm Inversion Code} (STiC) which assumes non-local thermodynamical equilibrium, plane-parallel geometry and includes \bdp{partial frequency redistribution (PRD)}. 
This inversion code (IC) recovers a depth-stratified model covering the photosphere, chromosphere and transition region from the inversion of spectropolarimetric observations.
We have used STiC to invert the \mgii intensity data observed with IRIS. However, on average, the time needed to recover such information is about 2 $CPU-hour/profile$.
Thus, to invert an IRIS map such as the one shown in Figure  \ref{fig:i_rp_pos_db_T_vlos_ne} -- which contains \bdp{$\approx220,000\ spectra$} --, takes $\approx440,000\ CPU-hours$. 

To reduce this computationally prohibitive task and allow for the inversion of large fields-of-view and time-series of data, we have created a {\it framework} based on the inversion results of \mgii profiles and several machine and deep learning techniques. This new approach allows us to reconstruct models (with similar accuracy as STiC) from any IRIS dataset in a few minutes using a desktop machine. Accompanying this paper, we make this code publicly available.

In section \ref{sec:building} we describe the foundations of the new framework. In section \ref{sec:iris2} and \ref{sec:deep} we present how the novel inversion methods work. The first results and their validation are shown in Section \ref{sec:results}. Finally, we discuss the advantage and limitations of the framework in Section \ref{sec:conclusions}. 


\section{IRIS \mgii database}\label{sec:building}

We have created a database of \mgii profiles observed with IRIS using the {\it Representative 
Profiles} (RP) of 250 datasets of different solar features, such as: quiet Sun,
plage, sunspots, emerging flux regions, active regions, flares, coronal
holes and filaments. The RPs are obtained after applying a clustering 
technique (k-mean analysis, \citealt{Mac67, Ste57}) to the spectral profiles of 
\mgii from the selected datasets. 

\asdrev{Each dataset in the database is clustered in 60 RPs}, 
which we have inverted with STiC using the same inversion scheme for all RPs. The number of RPs was determined by hardware constraints. The inversion setup 
consists of two cycles. The first cycle considers 4 nodes\footnote{The number of {\it nodes} is the number of grid points (or degrees of freedom) 
in which an atmospheric parameter is allowed to vary.} in temperature, and 3 
nodes both in micro-turbulence ($v_{turb}$) and line-of-sight velocity 
($v_{los}$). The second cycle takes as input the output model from the 
first cycle, now using 7 nodes in temperature, and 4 nodes 
both in $v_{turb}$ and $v_{los}$. \asd{Each RP is inverted 3 times from a different set of initial parameters (randomization) in each cycle.}

For each (observed) RP we obtain a synthetic RP (RP@STIC), which is the best match found by the IC between the observed and synthetic profiles, and the corresponding {\it Representative Model Atmosphere} (RMA). A 
RMA \bdp{consists of} the depth-stratifications of temperature ($T$ in $K$),  \vlos in 
$cm/s$, $v_{turb}$ in $cm/s$, gas pressure ($p_{gas}$ in $dyn/cm^2$), mass-density 
($\rho$ in $g/cm^{-3}$), electron density ($n_{e}$ in $cm^{-3}$), column mass ($c_{mass}$ in 
$g/cm^{2}$), and height ($z$ in $cm$).

\subsection{Physical meaning of the RPs and RMAs}

A RP is the averaged profile of a cluster of profiles sharing the same shape as a function of wavelength. From a machine learning perspective, the intensity at any \bdp{wavelength} is a {\it feature}. Thus, a profile in the IRIS \mgii database is a {\it sample} with 473 features, the number of wavelength points in the profiles.
The k-mean clustering technique clusters these features independently. In our case, the features determine the shape of the profile. 

The shape of a spectral profile encodes information of the atmosphere from which the radiation originates\footnote{In this paper we do not consider polarimetric data or magnetic field, as IRIS observes only intensity. However, the methods presented are also valid for spectropolarimetric data.}
Locations with similar physical conditions shall share profiles with similar shape; a region in the  atmosphere with similar conditions is associated with a RP -- or a few RPs.

The left panel of the top row of Figure \ref{fig:i_rp_pos_db_T_vlos_ne} shows the 
\bdp{IRIS} intensity map at the blue peak of Mg {\footnotesize II} $k$ line ($k_{2V}$ spectral feature) for NOAA AR 12480. In the central panel, 
the spatial distribution of the \bdp{corresponding} \mgii RPs is shown. There, we can appreciate how the RPs are distributed in coherent patches in the spatio-temporal \bdp{(since the raster scan takes time)} domain. 
The second row of Figure \ref{fig:i_rp_pos_db_T_vlos_ne} shows $T$, $v_{los}$, and $log(n_e)$ recovered from the inversion of the RPs, i.e. \asd{in} the RMAs of that dataset. 

We call this method inversion of RPs by STiC or $\mathbf{RPs@STiC}$. Because we are inverting a reduced number of profiles (the RPs), this method can provide valuable information of the physical conditions in the IRIS field of view (FoV) within a few CPU hours.

The \rps\ is a good method to recover information on spatially coherently averaged areas, although there is a loss of 
spatial information. 
Moreover, \asdrev{a few poorly fitted RPs may affect a large region}.
Thus, in the $v_{los}$ shown in Figure \ref{fig:i_rp_pos_db_T_vlos_ne}, the patches associated \bdp{with} the border between plage and quiet Sun show suspicious values. A close inspection \bdp{of} the quality of the fit of those RPs confirms that their match is not good. To avoid these flaws we have developed two more sophisticated methods. 

\subsection{Building the database}

We have considered most of the main solar features observed in the photosphere and chromosphere. In addition, \asdrev{the employed} data sets were selected considering position on the solar disk, exposure time, and IRIS observing modes: dense (0.33\arcsec steps) or sparse (1\arcsec) raster, and sit-and-stare. 
The location of all datasets included in the database \asdrev{is} indicated in Figure~\ref{fig:i_rp_pos_db_T_vlos_ne}.

The database consists of three elements: 15,000 observed RPs, the corresponding 15,000 synthetic RPs (from the inversion of the RPs), and the corresponding 15,000 model atmospheres. Because we have a large number of RPs, both the synthetic \asdrev{RPs} and RMAs represent \asdrev{the variety of typical solar conditions} quite well.

Our database is constructed from observations that are sensitive to the upper photosphere and chromosphere. Therefore, the IRIS \mgii database may also be useful beyond the direct purpose of this paper. For instance, theoretical models and numerical simulations may find valuable observational constraints in our \bdp{database.}

\section{Inversion of IRIS \mgii lines based on inverted RPs}\label{sec:iris2}
\begin{figure*}[h!]
	\centering
	\includegraphics[scale=.85, trim= 45 0 35 0, clip]{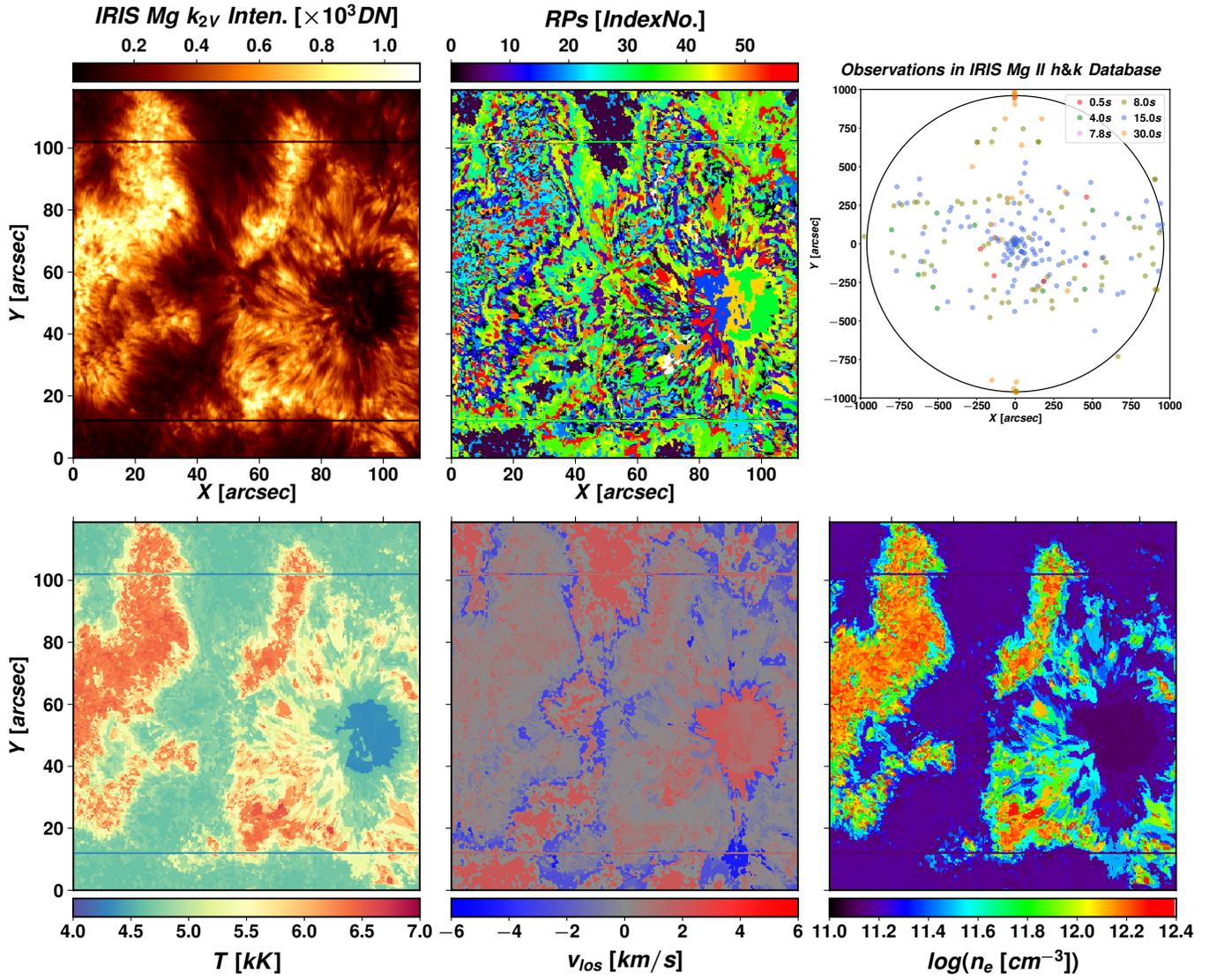}
	\caption{{\it Top}: Left: slit-reconstructed intensity map of NOAA AR 12480 observed by IRIS at Mg 
	{\footnotesize II} $k_{2V}$. Center: location of the representative profiles 
	(RPs). 
	Right: location on the solar disk of the IRIS \mgii database observations and their  (color-encoded) exposure time. {\it Bottom}: from 
	left to right, $T$,  $v_{los}$, and electron density ($log(n_e)$) evaluated at $log(\tau) = -4$.
	} 
	\label{fig:i_rp_pos_db_T_vlos_ne}
\end{figure*}

For any pixel of a given observation, e.g., the one shown in Figure 
\ref{fig:i_rp_pos_db_T_vlos_ne}, we look for the closest synthetic profile obtained by the \bdp{StiC inversion of the RP} in the IRIS \mgii database ($I_{i}^{syn\  RP@STiC}$). To determine the closest profile we use the same loss function as in STiC,
\begin{equation}\label{eq:chi2}
\chi^{2} = \frac{1}{\nu}\sum_{i=0}^{q}{(I(\lambda_i)^{obs} - I(\lambda_i, \mathbf{M})^{syn\  RP@STiC})^{2}\frac{w_{i}^2}{\sigma_{i}^2}} 
\end{equation}

with $i = 0,..., q$ the sampled \bdp{wavelengths},
$w_{i}$ \bdp{their weights}, $\sigma_{i}$ the uncertainties of the observation (e.g. photon noise), and $\nu$ the \asd{number of observables, i.e.}\bdp{, the spectral samples}. 
A low value of $\chi^{2}$ tells us whether the fit 
between the observed 
($I_{obs}$) and synthetic profiles $I_{i}^{syn\  RP@STiC}$ is \bdp{good.}
We explicitly denote the dependency of the synthetic RP on the parameters of the model atmosphere ($\mathbf{M}$).   
Once the code has found the best match between the observed profile and \asd{a synthetic RP} in the IRIS database, it associates the corresponding RMA of that (closest) synthetic RP to the pixel in our observation. 
For large datasets, this look-up table process may take a few minutes on a desktop machine. Then, the code provides a $\chi^{2}$-map \bdp{(to indicate the goodness of the match between the observed and synthetic profiles in the database)}, the output model atmosphere, and the associated uncertainty of each variable of the model\footnote{Currently, only uncertainties for $T$, $v_{los}$, $v_{turb}$, and $n_{e}$ are in the \irissq\  database.}.

The uncertainty of a physical quantity $p$ is calculated following Eq. 42 in \cite{delToroIniesta16}:
\begin{equation}
\sigma^{2}_p = \frac{2}{nm+r}\frac{\sum_{i=1}^{q}{\left[ I(\lambda_i)^{obs} - I(\lambda_i; \mathbf{M})^{syn\  RP@STiC})^{2}\right] \frac{w_{i}^2}{\sigma_{i}^2}}}{\sum_{i=1}^{q}R^2_{p}(\lambda_{i})\frac{w_{i}^2}{\sigma_{i}^2}}\label{Eq:unc}
\end{equation}
\bdp{with} $m$ the number of physical quantities in the model $\mathbf{M}$ evaluated in $n$ grid points along the solar atmosphere, $r$ the number of physical quantities considered constant along that atmosphere, and $R_{p}$ the {\it Response Function} (RF) of a Stokes parameter to the physical quantity $p$ \citep{Mein71,LandiDegl'Innocenti79,RuizCobo92}. 
The RF \bdp{provides the sensitivity of a wavelength sample in a Stokes profile} to (changes of) a physical quantity.  Thus, we use expressions like: {\it 	``The core of 
the \mgii lines is sensitive to the $T$ in optical depths around $log(\tau)\footnote{We use $log(\tau)$ to refer to  $log_{10}(\tau_{5000}$).} = -5$, while the wings are sensitive to $T$ in $-5 < log(\tau) < -1$''}. 

We note that the inversion code does not operate over every grid point of the atmosphere, but over the nodes (each of them usually affect several grid points simultaneously). Therefore, in the latter case the RF is usually larger as a larger section of the atmosphere is perturbed per node and the uncertainty becomes lower than the estimates obtained by perturbing each grid point.



We have named this new tool the {\it {\bf IRIS I}nversion based on {\bf R}epresentative Profiles {\bf I}nverted by {\bf S}TiC} or just {\bf IRIS squared} ($\mathbf{IRIS^{2}}$).


\irissq\ relies on two fundamental concepts: i) 
the relationship between the synthetic RPs and the RMAs, given by the inversion of the observed RPs by STiC, and ii) \bdp{because
the IRIS database covers a large variety of solar features, the RPs and corresponding RMAs are a meaningful representation \asdrev{of the variety found in} the chromosphere and upper photosphere}. 


\section{Inversion of IRIS \mgii lines using deep learning}\label{sec:deep}

Since the IRIS \mgii database \bdp{includes} synthetic profiles of the RPs and the corresponding RMAs, we have trained several deep neural networks (DNN) to reproduce this relationship. In deep learning jargon, a synthetic RP is the {\it input layer}, \bdp{i.e., the intensities of the synthetic RP at the sampled \bdp{wavelengths} are the {\it input nodes}} (473). The corresponding RMA is the {\it output layer}, \bdp{i.e., the values of a physical quantities along the atmosphere (39) are the {\it output nodes}}. Once the DNN is trained, we are\bdp{, in principle,} able to {\it predict} the physical quantity \asdrev{through} the atmosphere for a given IRIS \mgii profile.   

We have considered $T$, $v_{los}$, $v_{turb}$, and $n_{e}$ as independent variables with respect to the corresponding synthetic RP. That means, we have trained a DNN for each of these physical quantities. The DNNs have different topologies (number of hidden layers and nodes), loss functions, and dropout parameters (to avoid overfitting). All the DNNs we have built use a rectified linear unit (ReLU) as activation functions, and we use 80\% of the IRIS \mgii database 
as a \bdp{training set}, and the remaining as a test set. Similarly, we have trained the uncertainties (along the atmosphere) of these physical quantities. 

We have named this method {\bf deep IRIS squared} or $\mathbf{deepIRIS^2}$.  \asdrev{More detailed information about the used DNNs will be given in a follow-up paper.}

\section{\bdp{Validation and Discussion}}\label{sec:results}
\asdrev{To validate \rps, \irissq, and \deepiris}, 
we have inverted, with STiC, every other pixel \bdp{of the IRIS \mgii observation of NOAA AR12480 on January 14, 2016}. We consider the STiC results as the ground truth, but we should note that the STiC results also depend on initialization and are not guaranteed to provide a global minimum of the loss function.

Some results of using \rps\ method are shown in the bottom of Figure \ref{fig:i_rp_pos_db_T_vlos_ne}. The first row of Figure \ref{fig:T_uncert_rel} shows $T$ at \asdrev{$\log(\tau) = -4$} as result of the inversion using STiC (left), \irissq\ (center), and \deepiris\ (right).
Figure \ref{fig:vlos_vturb_ne} shows $v_{los}$ (top), $v_{turb}$ (middle), and $log(n_e)$ (bottom). 
Movies showing the variation of these parameters as a function of depth in the atmosphere are available in the electronic version.

\subsection{Discussion: the reliable uncertainty range}
 
\bdp{One}
question we should answer to validate our results is: {for a physical quantity at a given optical depth $p(\tau)$, \bdp{how large} is the unsigned difference between the value recovered by our method and the one obtained by STiC \bdp{compared to}
the uncertainty estimated using Eq.~\ref{Eq:unc}}? We define the {\it Uncertainty Multiplication Factor}  (UMF) as:
\begin{equation}
UMF_p\ (\tau) = \frac{|p^{method}(\tau)  - p^{STiC}(\tau)|}{\sigma_p^{STiC}(\tau)} \label{Eq:condition}
\end{equation}

\begin{figure*}[t!] 
	\centering
	\includegraphics[width=\textwidth]{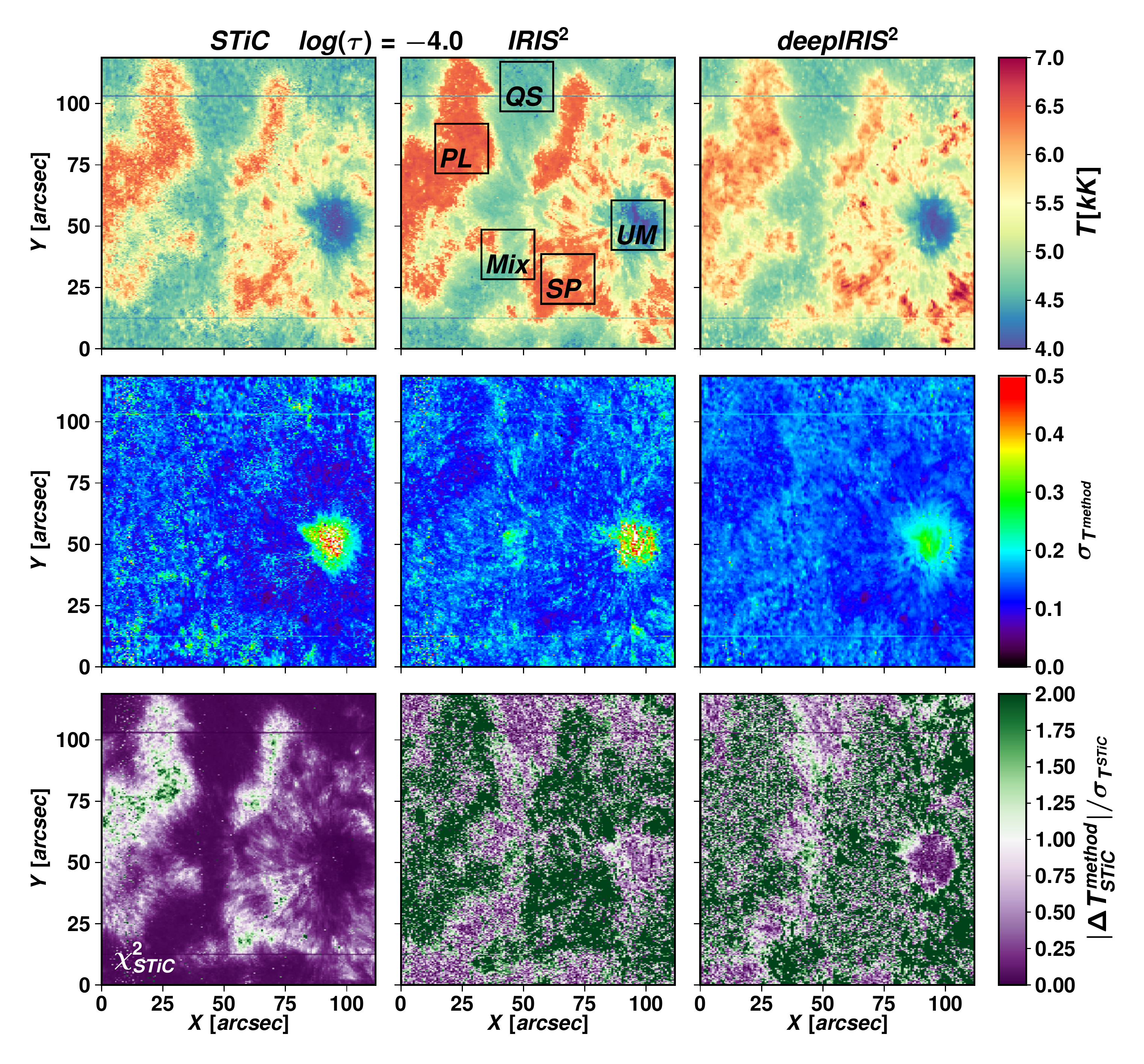}
	\caption{{\it Top}: First row: $T$ for NOAA AR 12480 at $log(\tau) = -4$ provided by STiC (left), \irissq\ (center), and \deepiris\ (right). {\it Middle}: uncertainties for these methods. {\it Bottom} Left: $\chi^2$-map associated to STiC results. Center and right: uncertainty multiplication factor ($UMF_T$)  for \irissq\ and \deepiris\ methods. Regions of interest (RoIs) are marked with squares in the center panel of the top row.
	\label{fig:T_uncert_rel}}
\end{figure*}

We have selected 5 regions of interest (RoIs) of $21''\times21''$ in the FoV: plage (PL), quiet Sun (QS), umbra (UM), superpenumbra (SP), and a mix of regions (Mix, see Figure \ref{fig:T_uncert_rel}) \asd{to help us in the interpretation of the results}. 
The full FoV (FFoV) is also evaluated.

The uncertainty maps ($\sigma$-maps) of $T$ obtained by STiC, \irissq\, and \deepiris\ are shown in the second row of Figure \ref{fig:T_uncert_rel}. The first panel of the third row shows the $\chi^2$ map for STiC.  The other panels in the third row show the \bdp{$UMF$} 
for $T$ using the \irissq\ (center) and \deepiris\ (right) methods. We have intentionally selected the optical depth $log(\tau) = -4$ because it \bdp{illustrates} several \bdp{key issues}.

\begin{figure*}[] 
\centering
\includegraphics[width=\textwidth]{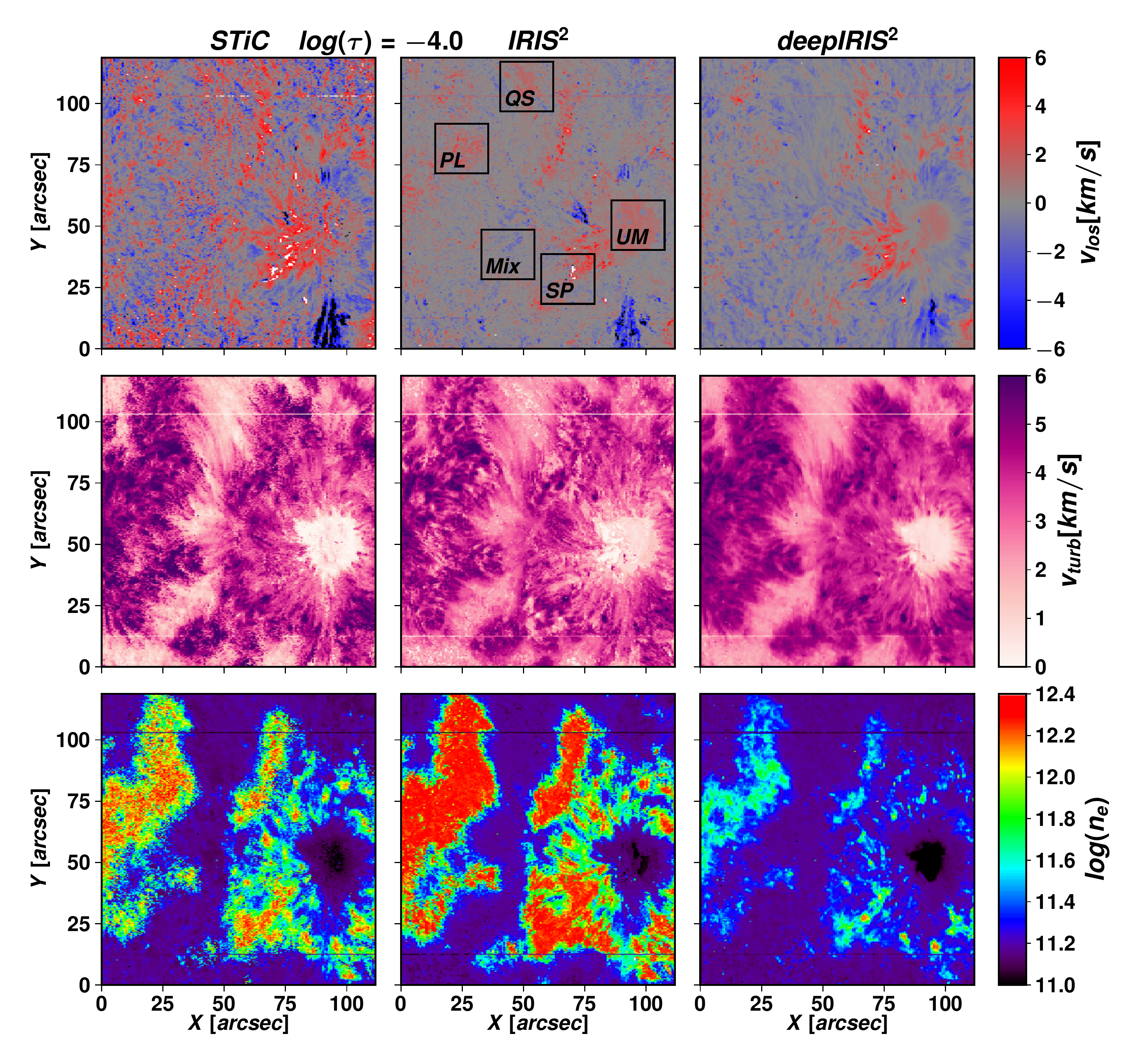} 
\caption{Some thermodynamic quantities recovered by STiC (left), \irissq\ (center), and \deepiris\ (right) at $log(\tau)=-4$: $v_{los}$ (top), $v_{turb}$ (middle), and $log(n_e)$ (bottom).
\label{fig:vlos_vturb_ne}}
\end{figure*}

Some regions in the $\sigma$-maps are better
 (lower values) for \irissq\ and \deepiris\ than for STiC\footnote{The values in the $\sigma$-maps for  \rps\ (not shown in this letter) are close to those for STiC.}. This is a direct consequence of the better fit obtained by \irissq\ compared to STiC, as mentioned above.
The $\chi^2$ map shown in the bottom-left panel of Figure \ref{fig:T_uncert_rel} is normalized in such a way that: \bdp{the fit is bad in regions where $\chi^2 \gg 1$, indicating that $p(\tau)$ is likely suffering from large uncertainties or may be wrong}; \bdp{the fits are better/good in regions with $\chi^2$ of order 1 or less.} 
In the \umf-maps (bottom center and right panels of Figure \ref{fig:T_uncert_rel}) regions with \umf $\approx 1$ have values of $p(\tau)$ as \bdp{"accurate" as STiC}, or even better if \umf $\ll 1$. \asdrev{Our example at $log(\tau)=-4$ indicates that} \bdp{care should be taken when interpreting the results in plage and the superpenumbra}. 

\begin{figure*}
	\centering
    \includegraphics[width=\textwidth, trim = 20 200 20 0, clip]{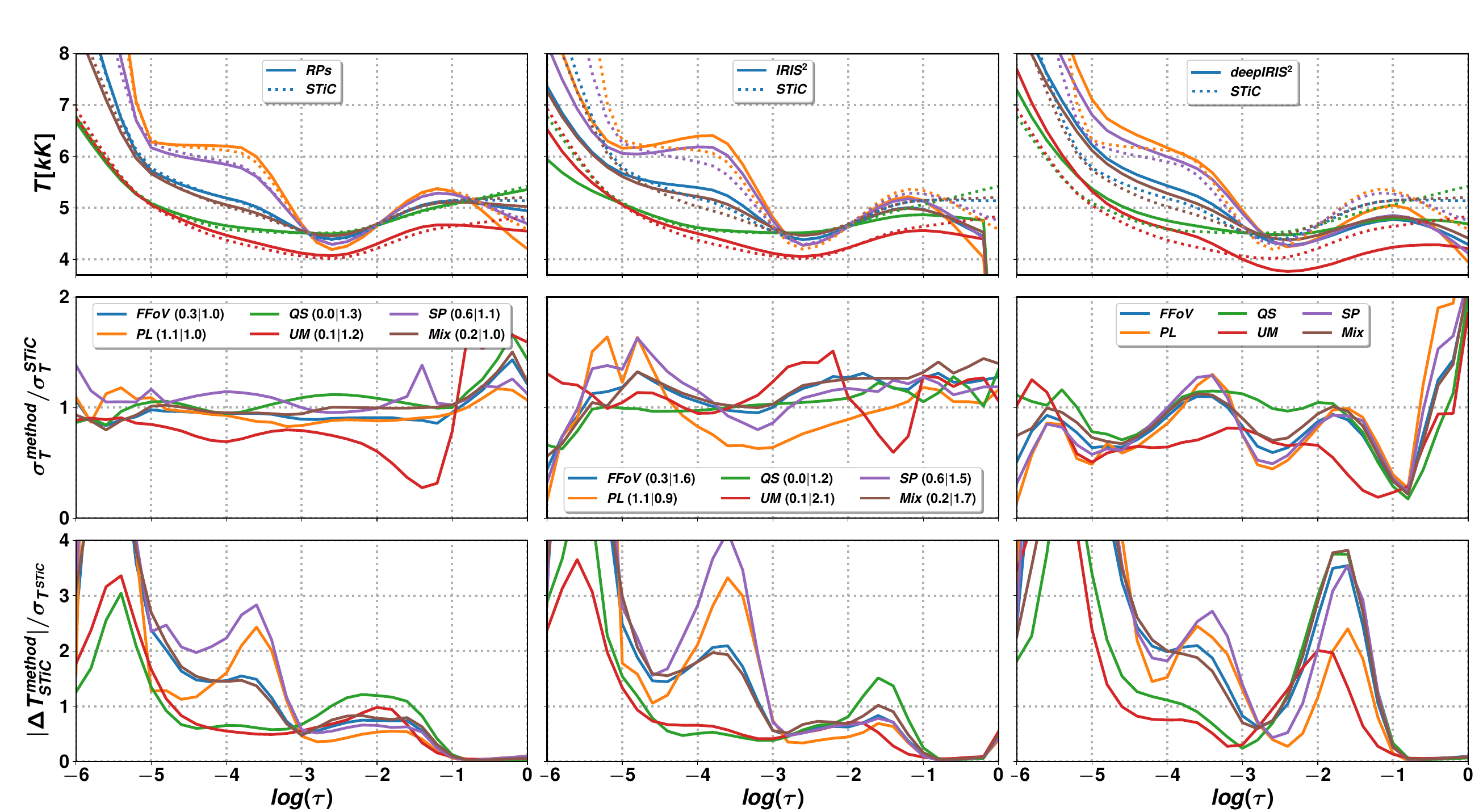}
	\caption{For the full FoV (FFOV, blue line) and the RoIs framed in Figure \ref{fig:T_uncert_rel}: {\it Top}: $T$ along the optical depth in the atmosphere ($log(\tau)$) for STiC (dashed line), \rps\ (left), \irissq\ (center), and \deepiris\ (right). {\it Middle:} ratio between the uncertainty of the new methods with respect to the uncertainty of STiC. {\ Bottom:} uncertainty multiplication factor ($UMF_T$) for the new methods (see Eq. \ref{Eq:condition})}.
	\label{fig:rel_uncert_joint_T_3x3}
	\vspace{3em}
\end{figure*}

Figure \ref{fig:rel_uncert_joint_T_3x3} shows the behavior of $T^{method}$ and $T^{STiC}$ (in thick and dashed line respectively in the first row), the ratio between the uncertainties ($\sigma^{method}_T/\sigma^{STiC}_T$, second row), and the \umf\ (third row) for the proposed methods.
The numbers next to the RoIs labels in the legend are the spatially-averaged normalized $\chi^2_{STiC}$ and the ratio $\chi^2_{method}/\chi^2_{STiC}$. 



The uncertainties ratios show that mostly all the methods \bdp{show} the same uncertainty as STiC. 
\asdrev{In some regions the other methods give better results, in other cases the opposite is true}, but on average (blue line) the behavior is very similar, or clearly better.

The \umf\ for \rps\ and \irissq\ is  $\lesssim 1$ in all RoIs for   $-3.5 < log(\tau) < -1$. For plage, the UMF for $ -4.5 < log(\tau) < -3$ is $\lesssim 1$,  and $\chi^2_{method} \le \chi^2_{STiC}$, therefore the values provided by these methods are more accurate than the ones \bdp{from} STiC. The situation is the opposite for the superpenumbra. All in all, the values provided by \rps\ and \irissq\ are mostly valid where \mgii lines are sensitive to $T$, i.e. $-5 < log(\tau) < -1$. For \deepiris,  we have to be cautious, even when the uncertainties ratio is $\lesssim 1$ in all the RoIs, the difference in $T$ is noticeable in some of the ROIs.

\begin{figure*}[h!]
	\centering
	\vspace{-2cm}
    \includegraphics[width=\textwidth, trim = 20 200 20 0, clip]{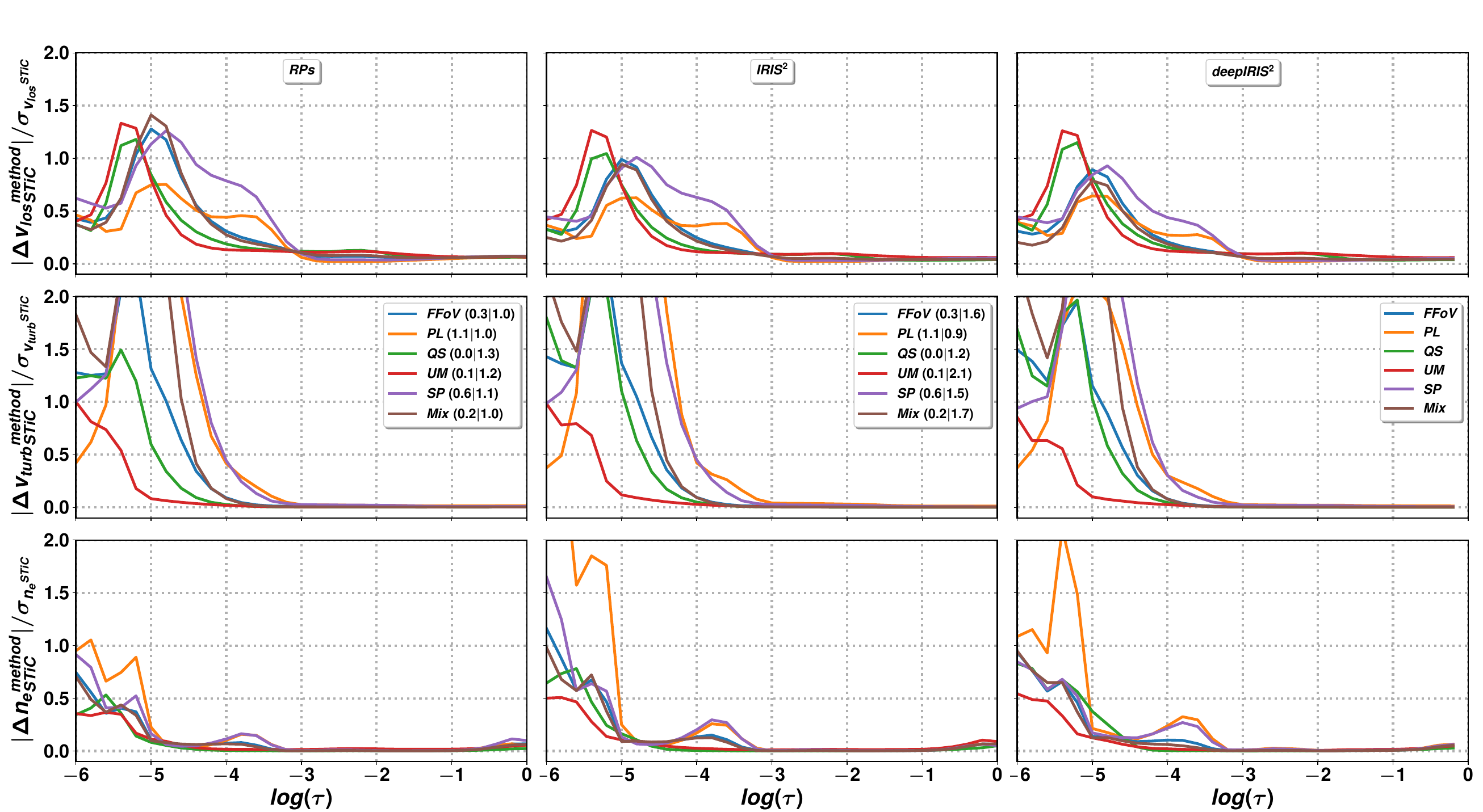}
	\caption{Uncertainty multiplication factor ($UMF_p$, see Eq. \ref{Eq:condition}) for $v_{los}$ (top), $v_{turb}$ (center), and $n_e$ (right) for the new methods for the full FoV (FFOV, blue line) and the RoIs framed in Figure \ref{fig:T_uncert_rel}.}
	\label{fig:rel_uncert_joint_vlos_vturb_ne}
\end{figure*}

Figure \ref{fig:rel_uncert_joint_vlos_vturb_ne} shows the \umf\ for $v_{los}$, $v_{turb}$, and $log(n_e)$. All the methods have a similar behavior in all the RoIs for $v_{los}$ and $v_{turb}$, but there are differences for $log(n_e)$ in plage regions for the \irissq\ and \deepiris methods. \bdp{We note that that while substantial differences of physical parameters between different methods can occur, these differences may often be smaller than the intrinsic uncertainty}.  

\asdrev{In summary, any of the proposed methods provides these values as well ast STiC -- or even better} --, 
within the intrinsic uncertainties. However, when we interpret a physical quantity and its uncertainty provided by our methods we should be specially cautious in: i)  regions showing large values in the $\chi^2$-map, i.e. where the fit between the observed profiles and the synthetic profiles is not good, and ii) those optical depths where \mgii lines are less sensitive to variations of a physical parameter. Under those conditions, the uncertainty will be larger, e.g., 2 or 3 times larger than the uncertainties provided in the database (as the UMF values in Fig. \ref{fig:T_uncert_rel} and \ref{fig:vlos_vturb_ne} suggest).     

\section{Conclusions}\label{sec:conclusions}

We have \bdp{created} and evaluated three novel methods to rapidly obtain the atmospheric physical quantities in the chromosphere and upper photosphere \asdrev{from the profiles of the IRIS \mgii lines}. The methods presented are valid for any spectro(polarimetric) data as far as they \bdp{can be inverted} by a traditional inversion code. We note that \irissq\ can be used for any IRIS observation that includes \ion{Mg}{2} h or \ion{Mg}{2} k (or both) lines. 

We summarize the main \bdp{advantages and disadvantages} for the three methods:

\begin{itemize}
	{\item \rps: on average, it is the closest to STiC. However, we lose spatial information. This can be minimized by using a much larger number of RPs for each dataset. It stills requires a proper inversion, which takes \bdp{hundreds} of CPU-hours \asdrev{(e.g. 320 CPU-hours for 160 RPs)}.}
	{\item \irissq: it offers results as good as STiC on average, being slightly better or worse than the latter in some solar features. The spatial information is almost as good as the original, although some regions may show little variation if the profiles are associated with the same RP in the database. That can be minimized by including a larger variety of profiles in the database. This method is $10^5-10^6$ times faster than STiC.}
	{\item \deepiris: it predicts values of $v_{vlos}$, $v_{turb}$, and $n_e$, as good as the ones obtained by STiC. The predicted values for $T$ are \bdp{not as good but} acceptable. A more complex DNN architecture and larger training and test datasets can improve this. The results do not lose spatial information and they look spatially smooth. It is $\approx 10^6$ times faster than STiC.}
\end{itemize}
    
As a result of our investigation, we conclude that \irissq\ is currently the fastest, easy-to-use method to recover reliable information from the chromosphere and photosphere from IRIS \mgii data. While we are improving these methods with a new database that includes 160 RPs per dataset, as well as more observations, we note that the current versions of IRIS \mgii lines database,  \irissq (both in IDL and Python), and \deepiris\ are available to the community at \href{http://iris.lmsal.com/analysis.html}{http://iris.lmsal.com/analysis.html}. We expect that our database can be applied to a wide variety of investigations that use IRIS data.  

\longacknowledgment{}

\end{document}